\documentclass[runningheads]{llncs}
\usepackage[T1]{fontenc}
\usepackage{graphicx}
\usepackage{amsmath,amssymb,amsfonts}
\usepackage{graphicx}
\usepackage{textcomp}
\usepackage{multirow}
\usepackage{tikz}

\usepackage{ctable}
\usepackage{amsmath,amssymb}
\usepackage{graphicx}
\usepackage{subcaption}
\usepackage{hyperref}
\usepackage{makecell}
\usepackage{cleveref}
\usepackage[commandnameprefix=ifneeded,final]{changes}

\usepackage{booktabs}
\usepackage{arydshln}

\begin{document}
\title{Myocardial Strain Drift Correction in Deep Learning Based Ultrasound Tracking}
\titlerunning{Myocardial Strain Drift Correction in Deep Learning Tracking}
%
\author{Thierry~Judge\inst{1,2}, Nicolas~Duchateau\inst{2,3}, Andreas~Østvik\inst{4,5,6}, Havard~Dalen\inst{4,5,7}, Bjørnar~Grenne\inst{4,5}, Pierre-Yves Courand\inst{2,8}, Lasse~Lovstakken\inst{4}, Pierre-Marc~Jodoin\inst{1}, and~Olivier~Bernard\inst{2,3}}
\authorrunning{T. Judge et al.}
%
\institute{
Dept. of Computer Science, University of Sherbrooke, Sherbrooke,  Canada. \and 
INSA, Université Lyon 1, CNRS UMR 5220, Inserm U1206, CREATIS, Villeurbanne, France. \and 
Institut Universitaire de France (IUF). \and 
Dept. of Circulation and Medical Imaging, NTNU, Trondheim, Norway \and 
Dept. of Cardiology and Cardiothoracic Surgery, St. Olavs Hospital, Trondheim, Norway \and 
Dept. of Health Research, SINTEF Digital, Trondheim, Norway \and 
Dept. of Medicine, Levanger Hospital, Nord-Trøndelag Hospital Trust, Levanger, Norway \and 
Cardiology Dept., Hôpital Croix-Rousse, Hospices Civils de Lyon, Lyon, France, and the Cardiology Dept., Hôpital Lyon Sud, Hospices Civils de Lyon, Lyon, France.
}
%
\maketitle              
\begin{abstract}

Myocardial strain from echocardiography is a key biomarker for cardiac function. Recent deep learning methods show strong performance for myocardial motion tracking but often lack physiological constraints, leading to temporal drift across the cardiac cycle. Consequently, tracked points may not return to their relative initial positions at the end of each cardiac cycle, producing inaccurate strain estimates and even divergence in some cases. We propose a deep learning framework that compensates for drift during myocardial tracking. We extend a state-of-the-art echocardiographic tracking method (TAS-Net) with persistent memory tokens that share information across sliding windows over full cardiac cycles. A teacher-student fine-tuning strategy on real echocardiographic data then enforces physiologically consistent cyclic motion while preserving tracking accuracy. Experiments show reduced global and regional strain drift, improved agreement with clinical references, and better test-retest reproducibility, supporting more reliable myocardial strain estimation in clinical practice. Code is available at \url{https://github.com/ThierryJudge/cardiac-motion}.

\end{abstract}
\section{Introduction}
Myocardial strain, which quantifies the deformation of the cardiac wall across the cycle, is highly informative for the assessment of cardiac dysfunction \cite{Mihos:Circ:2025}. In speckle tracking echocardiography (STE), it can be quantified locally after tracking speckle patterns, this technique being the clinical standard available in several clinical software packages.
However, such software packages may not be fully transparent regarding the post-processing involved (hence the lack of consensus about computational standards \cite{Voigt:EHJCI:2015}) and are not freely available. Besides, their performance may be limited on lower-quality images as often encountered in clinical routine. 

In contrast, deep learning methods derived from optical flow estimation have the potential to overcome these limitations. Such methods are either trained on a large dataset of labels acquired with speckle tracking~\cite{azadEchoTrackerAdvancingMyocardial2024,chernyshovLowComplexityPoint2025}, or a simulated dataset generated with known motion~\cite{evainMotionEstimationDeep2022,taskenEstimationSegmentalLongitudinal2025,judge2026deeplearningstrainestimation}. In both cases, while the deep learning-based methods can achieve competitive results with speckle tracking, they often lack physiological constraints in their training. One major constraint is drift, which refers to the accumulation of errors across the sequence, resulting in different (and sometimes divergent) strain values from one cycle to another. Without drift, strain should theoretically be zero at every end-diastolic (ED) instant, for purely cyclic sequences. This phenomenon is most evident when strain values, such as Global (GLS) or Regional Longitudinal Strain (RLS), do not return to zero at the end of the cycle, reflecting a mismatch between the final and initial relative positions of the tracked points. Despite its importance, drift is not explicitly addressed in most 
tracking 
approaches. Existing 
deep learning 
models are largely adapted from general-purpose tracking architectures developed for natural images and do not enforce cyclic consistency. Current commercial solutions typically rely on a posteriori drift correction, which is suboptimal as it does not influence the tracking process itself.

In this work, we propose a deep learning framework for myocardial tracking that explicitly accounts for drift and enforces physiologically consistent cyclic motion. Our approach extends the TAS-Net model with mechanisms to capture long-range temporal dependencies across the full cardiac cycle. In addition, we introduce a teacher-student fine-tuning strategy on real echocardiographic data, combining a novel drift loss with shape-preserving and temporal regularization terms. By integrating these constraints directly into the learning process, our method improves the reliability of myocardial strain estimation and reduces drift without compromising tracking accuracy.

\section{Method}

\begin{figure*}[tp]
\centering
\includegraphics[width=\textwidth]{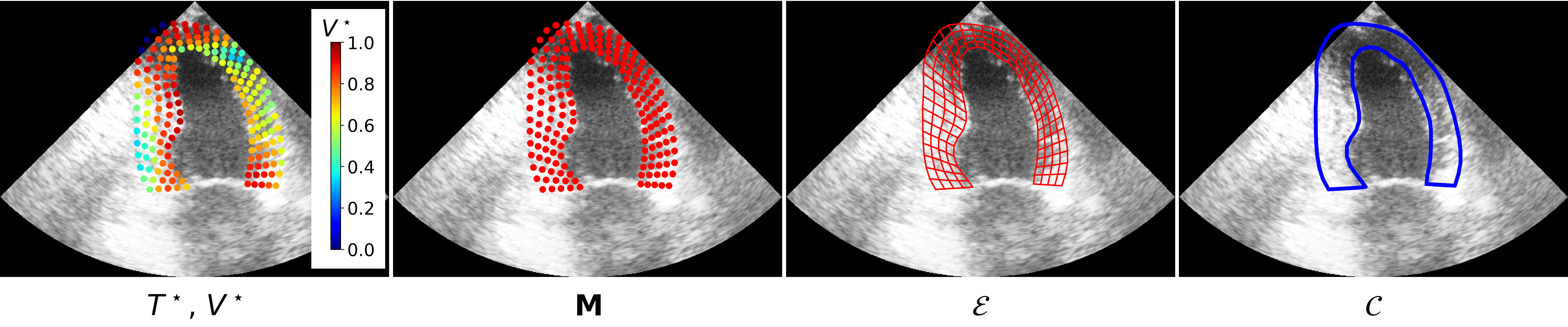}
\caption{Training elements of TAS-Net-D for a given frame: teacher tracks and visibility $(T^\star, V^\star)$, predicted mesh $\mathbf{M}$, edge set $\mathcal{E}$, and contour $\mathcal{C}$. }
\label{fig:method}
\end{figure*}

We aim 
to estimate myocardial longitudinal strain from a video $\mathbf{V} \in [0,1]^{T \times H \times W}$, composed of $T$ frames of height $H$ and width $W$, and a set of query points along the myocardium. 
These $K$ points are organised in a grid of radial and longitudinal points (\cref{fig:method}), which we consider at a specific time $t$ as \mbox{$M_t \in \mathbb{R}^{K\times 2}$}. 
The goal is to track these points through the video and to obtain a spatio-temporal mesh \mbox{$\mathbf{M} \in \mathbb{R}^{T \times K \times 2}$}.  With the centerline of this mesh, longitudinal strain can be computed with the usual equation: $\mbox{Strain} = \frac{l_t - l_{ED}}{l_{ED}}$, where $l$ is the length of the entire centerline in the case of the GLS or the length of a segment in the case of the RLS. In both cases, issues occur if the strain value at the second ED instant does not return to zero or, worse, increases over each cardiac cycle. Correcting drift requires the model to see the entire cardiac cycle. \textit{Offline} trackers such as EchoTracker~\cite{azadEchoTrackerAdvancingMyocardial2024}, MyoTracker~\cite{chernyshovLowComplexityPoint2025}, and CoTracker3~\cite{cotracker3} do this but perform poorly in our experiments. TAS-Net~\cite{judge2026deeplearningstrainestimation} gets better results, but its CoTracker1 architecture~\cite{karaevCoTrackerItBetter2025} uses sliding windows (\cref{fig:windows}~[left]) that block information flow between the beginning and end of the cycle. We therefore propose architectural modifications and a fine-tuning strategy to correct drift.

\subsection{Model architecture}

For tracking, 
we extend 
the TAS-Net~\cite{judge2026deeplearningstrainestimation} model, which uses the CoTracker1 architecture.
Since most GPUs cannot accommodate the processing of full echocardiographic videos due to memory constraints, TAS-Net uses sliding window computations. At first, the model computes features with a CNN encoder from which track features are sampled. For all time steps, each tracked point has a $D$-dimensional track feature, resulting in a vector $Q \in \mathbb{R}^{T_w\times K \times D}$ for a window of length $T_w$.  Each track feature is initialized with the feature sampled at the initial location of the query points in the query frame and copied to all other frames. The tracked points are initialized by copying the queries for all frames in the first window. Track features are used to compute correlation volumes and both the track features and coordinate predictions are refined with $I$ transformer iterations. 
All windows overlap by half the window length. 
Subsequent windows initialize the overlap from the previous prediction and copy the last overlap frame's positions to the remaining frames. TAS-Net modified the forward only sliding window with a bidirectional sliding window to allow initialization at the ES frame, where the myocardium is better seen on ultrasound videos. TAS-Net also added a mesh embedding to identify each point in the mesh. 

The sliding window architecture limits temporal context to frames within a single window, preventing information from propagating across the full sequence. We therefore propose two major architectural changes 
to TAS-Net. 

First, we partially invert the window and refinement iteration order (\cref{fig:windows}). Instead
of fully refining each window before advancing to the next, we refine each window for only $I_i$ inner iterations before sliding to the next window, with iterations matching the refinement iterations used in CoTracker1. Once all windows have been processed, we perform another outer refinement pass over the entire video. Repeating this procedure for $I_o$ outer iterations interleaves refinement across windows rather than completing each window independently. 
TAS-Net refines each window for $I$ iterations (4 during training and 6 during inference). 
To keep the computation comparable, in TAS-Net-D, we use $I_i = 2$ for both training and inference, and set $I_o = 2$ during training and $I_o = 3$ during inference, resulting in the same total number of refinement iterations as the baseline.

Second, as shown in \cref{fig:windows}~[Right], we introduce window-persistent tokens that carry information between windows. 
This allows the model to accumulate information over the full video, to reduce drift. 
We 
use 
a set of $K$ window-persistent tokens of dimension $D$ (i.e., $\in \mathbb{R}^{K \times D}$) that carry information across sliding windows. At each window, the transformer reads from these tokens via a cross-attention block inserted between the spatial and temporal attention layers. After each window refinement, the tokens are updated through gated cross-attention with the current appearance features.  As in CoTracker1, the coordinates and track features are detached after each window refinement. However, in TAS-Net-D, the persistent tokens are only detached at the end of each outer iteration, making them the sole pathway for gradient flow across windows.

\subsection{Losses}

\begin{figure*}[tp]
\centering
\includegraphics[width=\textwidth]{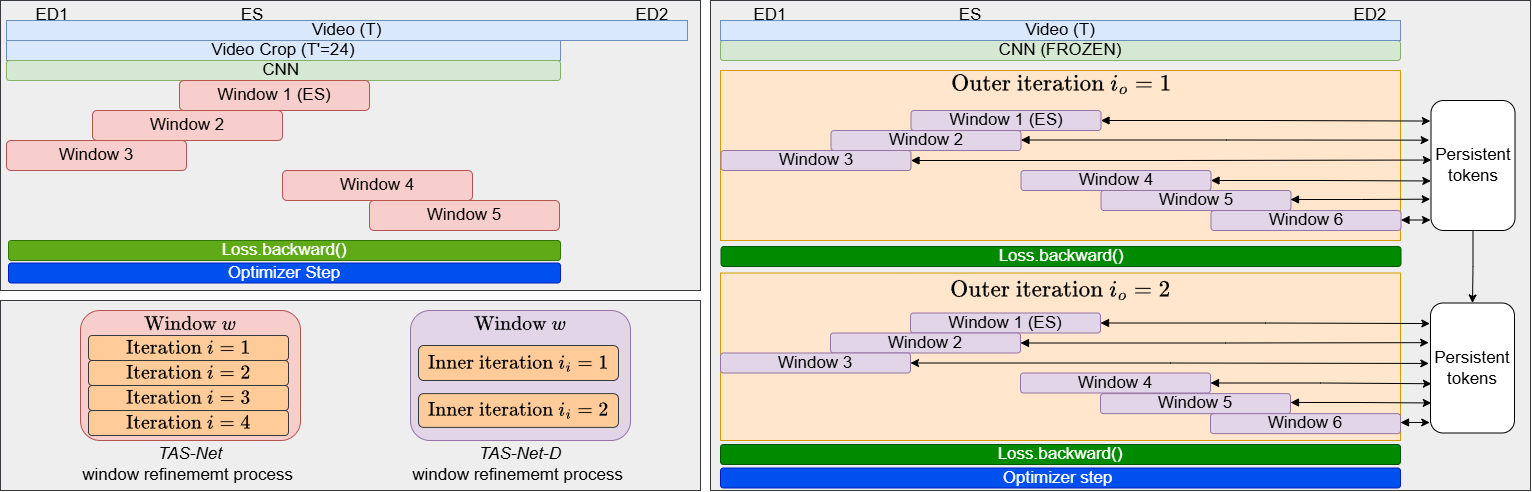}
\caption{
Video processing during training for TAS-Net [left] and TAS-Net-D [right]. Unlike TAS-Net's fully-refined windows, TAS-Net-D interleaves refinement across windows and introduces persistent tokens, enabling drift correction using information from the full sequence.
}
\label{fig:windows}
\end{figure*}

As described in the original paper~\cite{judge2026deeplearningstrainestimation}, TAS-Net is trained on simulated data that consists of highly realistic synthetic sequences reproducing existing real sequences with known speckle patterns, whose motion is controlled and stands as training ground truth. However, as this simulated data was generated by an iterative process in which TAS-Net generates the motion used to create simulated videos, drift can be present and compound over dataset iterations. It is therefore not practical to trivially add a loss function to correct drift. This loss could go against the simulated motion. We therefore propose a fine-tuning training process on real data using a pre-trained model (the original TAS-Net) as a teacher. 

Let's denote $\mathbf{M} \in \mathbb{R}^{T \times K \times 2}$ to define the predicted points of the model (c.f.~\cref{fig:method}). As mentioned before, during training, supervision is done over $I_o$ outer iterations and a set of windows $\mathcal{W}$ of length $T_w$ frames (see~\cref{fig:windows}). We therefore refer to the prediction at a given iteration $i$ and window $w$ as: $\mathbf{M}^{(i,w)} \in \mathbb{R}^{T_w \times K \times 2}$.  The teacher output contains both the point coordinates $T^\star\in\mathbb{R}^{T\times K\times 2}$, and a visibility parameter at each point $V^\star\in[0,1]^{T\times K}$. 

Instead of focusing on the longitudinal strain drift, we take a more general approach. We consider the length between each neighboring point in the mesh. We define $\mathcal{E}$ as the set of edges $e_{ab}$ connecting neighboring nodes $a$ and $b$ in the radial or longitudinal directions. For a given mesh $\mathbf{M}$, the length of each edge $e$ at  time $t$ is given by $\ell_{t,e} = ||\mathbf{M}_{t,a} - \mathbf{M}_{t,b}||. $
We use 
a combination of four losses to correct the drift while ensuring consistent predictions with the teacher. 

\noindent\textbf{Drift loss.}
The drift loss aims at minimizing the difference in edge lengths between two ED instants referred to as $t_1$ and $t_2$. As multiple windows can contain the ED indices, we average across windows that contain a given time $t$ ($t_1$ or $t_2$), $\bar{\mathbf{M}}^{(i)}_{t} = \frac{1}{|\mathcal{W}(t)|} \sum_{w \in \mathcal{W}(t)} \mathbf{M}_t^{(i, w)} \in \mathbb{R}^{K\times 2}$. 
We can compute the length of each edge at both $t_1$ and $t_2$ for iteration $i$ to get $\ell_{t_1,e}^{(i)}$ and $\ell_{t_2,e}^{(i)}$. We minimize the drift using the SmoothL1 loss defined in \cite{smoothL1}:


\begin{equation}
\mathcal{L}^{(i)}_{\mathrm{d}} =  \frac{1}{|\mathcal{E}|}\sum_{e\in\mathcal{E}}\operatorname{SmoothL1}\Big(\frac{\ell_{t_1,e}^{(i)}-\ell_{t_2,e}^{(i)}}{\text{sg}(\ell_{t_1,e}^{(i)})}\Big),
\end{equation}
%
where sg is the stop-gradient operator, which makes the normalization a fixed relative scale rather than a target the model can increase.

\noindent\textbf{Anchor losses.}
We use two anchor losses to tie the prediction to the teacher. The first ensures that the global mesh shape stays consistent with that predicted by the teacher, using a Chamfer loss~\cite{chamfer} applied to each window. We define the part of the prediction that makes up the contour (endocardium, epicardium and base) as $\mathcal{C}$, and $\mathcal{Q}^\star$ for the teacher. The loss is given by:
\begin{equation}
    \mathcal{L}_c^{(i)} = \frac{1}{|\mathcal{W}|} \frac{1}{T_w} \sum_{w \in \mathcal{W}} \sum_{t=1}^{T_w} \text{Chamfer}(\mathcal{C}^{(i,w)}_t, \mathcal{Q}^{\star(i,w)}_t).
\end{equation}
We also use a point-to-point loss between the teacher and student predictions.  To prioritize reliable supervision, each term is weighted by the visibility predicted by the teacher, concentrating the point-to-point loss on high-visibility points and leaving the drift correction loss free to dominate in low-visibility regions:
\begin{equation}
    \mathcal{L}^{(i)}_p = \frac{1}{|\mathcal{W}|}\frac{1}{T_w}\frac{1}{K} \sum_{w \in \mathcal{W}}  \sum_{t=1}^{T_w} \sum_{k=1}^K  V^{\star(w)}_{t,k} \cdot ||\mathbf{M}^{(i,w)}_{t,k} - T^{\star(w)}_{t,k} ||_1 
\end{equation}

\noindent\textbf{Regularization loss.}
As the drift loss 
only involves the two ED instants, the loss can be minimized by only updating the mesh at these instants, resulting in a spike in the strain curves. We therefore add a loss regularizing the third temporal derivative of the edge length to penalise rapid changes in acceleration. 
We can compute the edge lengths for each iteration and window, $\ell^{(i,w)}_{t,e}$,  and minimise
\begin{equation}
    \mathcal{L}_s^{(i)} = \frac{1}{|\mathcal{W}|} \frac{1}{T_w}  \frac{1}{|\mathcal{E}|}  \sum_{w \in \mathcal{W}} \sum_{t=1}^{T_w} \sum_{e \in \mathcal{E}} ||\ell^{(i,w)}_{t,e} - 3\ell^{(i,w)}_{t-1,e} + 3\ell^{(i,w)}_{t-2,e} - \ell^{(i,w)}_{t-3,e}||^2.
\end{equation}

\noindent\textbf{Training.}
The final loss, $\mathcal{L} = \sum_{i=0}^{I_o-1} \gamma^{\,I_o-1-i} \big(
  \lambda_{\mathrm{d}}\,\mathcal{L}^{(i)}_{\mathrm{d}}
  +\lambda_{\mathrm{s}}\,\mathcal{L}^{(i)}_{\mathrm{s}}
  +\lambda_{\mathrm{c}}\,\mathcal{L}^{(i)}_{\mathrm{c}}
  +\lambda_{\mathrm{p}}\,\mathcal{L}^{(i)}_{\mathrm{p}}\big)$, sums 
all losses over all outer iterations,  
%
where $\gamma=0.8$ is a discount factor to reduce the weight of earlier iterations. As we do not need motion reference for any of the loss terms, we train TAS-Net-D on real videos. We use a pre-trained TAS-Net model as the teacher and also to initialise the weights, except for token-related operations, which are initialised to have no effect on the computation before training. During training, 
freezing 
the CNN encoder
allows 
reducing the GPU memory and 
processing full video sequences. 
To accommodate large videos, we also accumulate the gradient of the loss between each outer iteration before taking an optimizer step after processing all the outer iterations.

\section{Experimental setup}
\begin{table}[tp]
\begin{tabular*}{\textwidth}{@{\extracolsep{\fill}}l ccc ccc ccc}

\toprule
\multirow{2}{*}{Method} & \multirow{2}{*}{\makecell{MTE\\(mm)}} & \multirow{2}{*}{\makecell{GLS drift\\(\%)}}& \multirow{2}{*}{\makecell{RLS drift\\(\%)}}& \multicolumn{3}{c}{Reference (\%)}& \multicolumn{3}{c}{Test-retest (\%)} \\\cmidrule(lr){5-7} \cmidrule(lr){8-10}

& & & & MAE  & $\mu$  & $\sigma$  & MAE  & $\mu$  & $\sigma$    \\
\midrule
Reference & - & 0.32 & 1.04 & - & - & - & 1.99 & -0.28 & 2.51 \\
$\text{EchoTracker}_\text{R}$ & 1.35 & 0.92 & 2.30 & 1.53 & 0.55 & 1.79 & 1.87 & 0.07 & 2.38 \\
\midrule
EchoTracker & 1.97\tiny{$\pm$0.06} & 2.30\tiny{$\pm$0.31} & 4.98\tiny{$\pm$0.36} & 3.40 & 3.25 & 2.68 & 2.63 & -0.63 & 3.34 \\
MyoTracker & 1.60\tiny{$\pm$0.04} & 1.06\tiny{$\pm$0.02} & \textbf{2.18}\tiny{$\pm$0.01} & \textbf{1.64} & \textbf{0.88} & 2.00 & 2.57 & -0.11 & 3.14 \\
CoTracker3 & 1.39\tiny{$\pm$0.01} & 1.10\tiny{$\pm$0.01} & 2.74\tiny{$\pm$0.05}& 1.77 & 1.43 & 1.74 & 2.21 & -0.20 & 2.73 \\
TAS-Net & 1.28\tiny{$\pm$0.00} & 1.20\tiny{$\pm$0.05} & 3.58\tiny{$\pm$0.05} & 1.84 & 1.68 & 1.54 & 1.92 & \textbf{-0.00} & 2.46 \\
    \hspace{1em}$\hookrightarrow$ Linear corr. & 1.34\tiny{$\pm$0.01} & 0.00\tiny{$\pm$0.00} & 0.00\tiny{$\pm$0.00} &1.90 & 1.80 & 1.49 & 1.89 & -0.16 & 2.40 \\
TAS-Net-D & \textbf{1.22}\tiny{$\pm$0.01} & \textbf{0.98}\tiny{$\pm$0.06} & 2.35\tiny{$\pm$0.07}  & 1.81 & 1.68 & \textbf{1.45} & \textbf{1.83} & -0.16 & \textbf{2.33} \\
\cdashline{1-10}
\textcolor{gray}{TAS-Net-D{\tiny$\lambda_s=0$}} &
\textcolor{gray}{1.24\tiny{$\pm$0.01}} & \textcolor{gray}{1.03\tiny{$\pm$0.03}} & \textcolor{gray}{2.64\tiny{$\pm$0.12}} &
\textcolor{gray}{2.03} & \textcolor{gray}{1.92} & \textcolor{gray}{1.50} &
\textcolor{gray}{1.78} & \textcolor{gray}{-0.15} & \textcolor{gray}{2.27} \\

\textcolor{gray}{TAS-Net-D{\tiny$\lambda_c=0$}} &
\textcolor{gray}{1.24\tiny{$\pm$0.00}} & \textcolor{gray}{1.09\tiny{$\pm$0.10}} & \textcolor{gray}{2.44\tiny{$\pm$0.13}} &
\textcolor{gray}{1.96} & \textcolor{gray}{1.85} & \textcolor{gray}{1.50} &
\textcolor{gray}{1.66} & \textcolor{gray}{-0.14} & \textcolor{gray}{2.17} \\

\textcolor{gray}{TAS-Net-D{\tiny$\lambda_p=0$}} &
\textcolor{gray}{1.26\tiny{$\pm$0.01}} & \textcolor{gray}{0.93\tiny{$\pm$0.02}} & \textcolor{gray}{1.95\tiny{$\pm$0.07}} &
\textcolor{gray}{2.15} & \textcolor{gray}{2.06} & \textcolor{gray}{1.45} &
\textcolor{gray}{1.67} & \textcolor{gray}{-0.16} & \textcolor{gray}{2.17} \\

\bottomrule
\end{tabular*}
\caption{Results on the HUNT dataset. \replaced{Mean and std. of three runs are reported for MTE and drift metrics. The median run according to Reference GLS MAE is reported for Bland-Altman statistics}{(three runs performed for each method; we report the median run according to Reference GLS MAE)}. 
\textbf{Bold}: best values for methods supervised with simulated data. 
$\text{EchoTracker}_\text{R}$ is supervised with STE-derived labels. Gray: ablation study for the different loss components.
}
\label{tab:HUNT}
\end{table}

We compare our method to other 
state-of-the-art 
trackers: EchoTracker~\cite{azadEchoTrackerAdvancingMyocardial2024}, MyoTracker~\cite{chernyshovLowComplexityPoint2025}, CoTracker3~\cite{cotracker3}, 
and 
the original TAS-Net~\cite{judge2026deeplearningstrainestimation}. We train the four models on the TAS-1K dataset~\cite{simu_IUS,judge2026deeplearningstrainestimation}
(1,478 simulated videos containing  $\sim 1$ cardiac cycle from the CAMUS~\cite{leclercDeepLearningSegmentation2019} and CARDINAL~\cite{lingExtractionVolumetricIndices2023} datasets). All methods were trained as described in their original papers. We trained CoTracker3 like TAS-Net, with random query initialization at training and ES initialization during inference and with mesh embedding. We also use a second version of EchoTracker, $\text{EchoTracker}_\text{R}$, which is trained on 6,490 real A2C, A3C, and A4C echocardiographic videos and released with the original paper. \added{We also apply an \textit{a posteriori} linear drift-correction to the TAS-Net
output. The conventional form distributes the drift across the cardiac cycle as
$\mathbf{M}^{\mathrm{corr}}_t = \mathbf{M}_t - \frac{t - t_1}{t_2 - t_1}(\mathbf{M}_{t_2} - \mathbf{M}_{t_1})$,
which holds $t_1$ fixed. Since TAS-Net is initialised at end-systole ($t_{\mathrm{ES}}$),
we instead anchor the correction at $t_{\mathrm{ES}}$ and split the drift symmetrically
over the first and second half of the cycle, so that the initialisation mesh is left
unchanged.
}

We trained TAS-Net-D on the REAL videos from the CAMUS and CARDINAL datasets corresponding to the original videos used to generate the TAS-1K dataset. We trained for 100,000 steps with the AdamW optimizer~\cite{adamw} and a one-cycle learning rate scheduler~\cite{onecycle} using an initial learning rate of $1 \times 10^{-4}$. \replaced{Loss weights were set heuristically to}{We used loss weights of} $\lambda_d = 30$, $\lambda_p=2$, $\lambda_s=1$ and $\lambda_c = 1$, in an attempt to equalize the drift loss's contribution with the anchor losses at the start of training.

The evaluation was conducted on the HUNT test--retest dataset~\cite{HUNT_SSHF}
consisting 
of 30 patients. For each subject, two independent echocardiographic acquisitions were performed in each of the three standard apical views (A2C, A3C, and A4C). Each recording was analyzed independently by a different clinical expert, enabling the assessment of inter-expert variability.
Reference annotations were generated semi-automatically by clinical experts using speckle tracking from the 
EchoPac commercial software, 
producing a sequence of $K_c$ (on average 73) points of size $T \times K_c \times 2$
along the myocardial centerline. 
This 
centerline is partitioned into six anatomical segments for RLS analysis. EchoPac scores each segment automatically using proprietary criteria. The expert may confirm or reject this score:
82\% of segments 
were approved
in the HUNT dataset. We derived reference GLS directly from the exported tracking points instead of EchoPac's own GLS output, since the latter depends on undisclosed proprietary computation.

We evaluated all methods against 
EchoPac. 
For all, tracking 
was initialised with the reference points at the ES frame, except for EchoTracker, which worked better with ED initialisation. 
We first computed the mean trajectory error (MTE) between the tracked points and the corresponding reference points for the accepted segments. We also evaluated the predicted peak GLS value against the reference value with mean absolute error (MAE) and with a Bland–Altman analysis. Finally, we computed average drift for GLS and RLS. 






\begin{figure*}[tp]
\centering
\includegraphics[width=\textwidth]{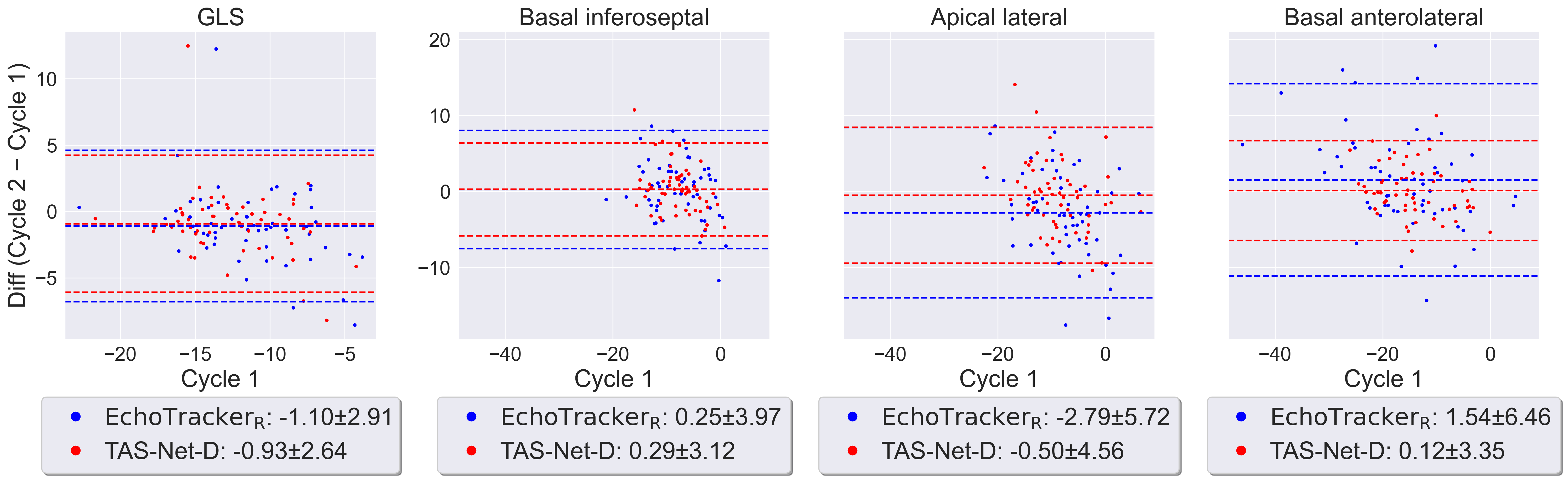}
\caption{Bland-Altman plot for GLS and RLS for CAMUS A4C multi-cycle videos. TAS-Net-D has a mean closer to zero and a smaller variance.}
\label{fig:ba}
\end{figure*}

\section{Results}


\Cref{tab:HUNT} shows results on the HUNT dataset. 
We observe that two models operating on full videos, MyoTracker and CoTracker3, exhibit low GLS and RLS drift and good GLS agreement with the reference. However, their MTE suggests poor alignment with the reference over the entire cycle, which is not represented in the GLS, which only considers two frames relative to each other. Moreover, their reproducibility in the test–retest setting is lower. \added{Applied to TAS-Net, the \textit{a posteriori} linear correction ({\em Linear corr.} in \Cref{tab:HUNT}) drives GLS and
RLS drift to zero by construction, but at the cost of a higher MTE and with no gain in reference agreement or test--retest
reproducibility, illustrating some limits of correcting drift after tracking.} TAS-Net-D slightly improves over TAS-Net for GLS agreement and is lower than other methods for MTE. Its GLS agreement with the reference is poorer than $\text{EchoTracker}_\text{R}$, which was trained on EchoPac-derived labels. In the test–retest setting, TAS-Net-D achieves the best MAE and variance ($\sigma$), even surpassing the reference. The ablation study shows that by removing some of the loss components, the drift can be minimized further and the agreement can be reduced; however, this is at the expense of GLS error with respect to the reference, justifying the three components. 



\begin{figure*}[tp]
\centering
\includegraphics[width=0.85\textwidth]{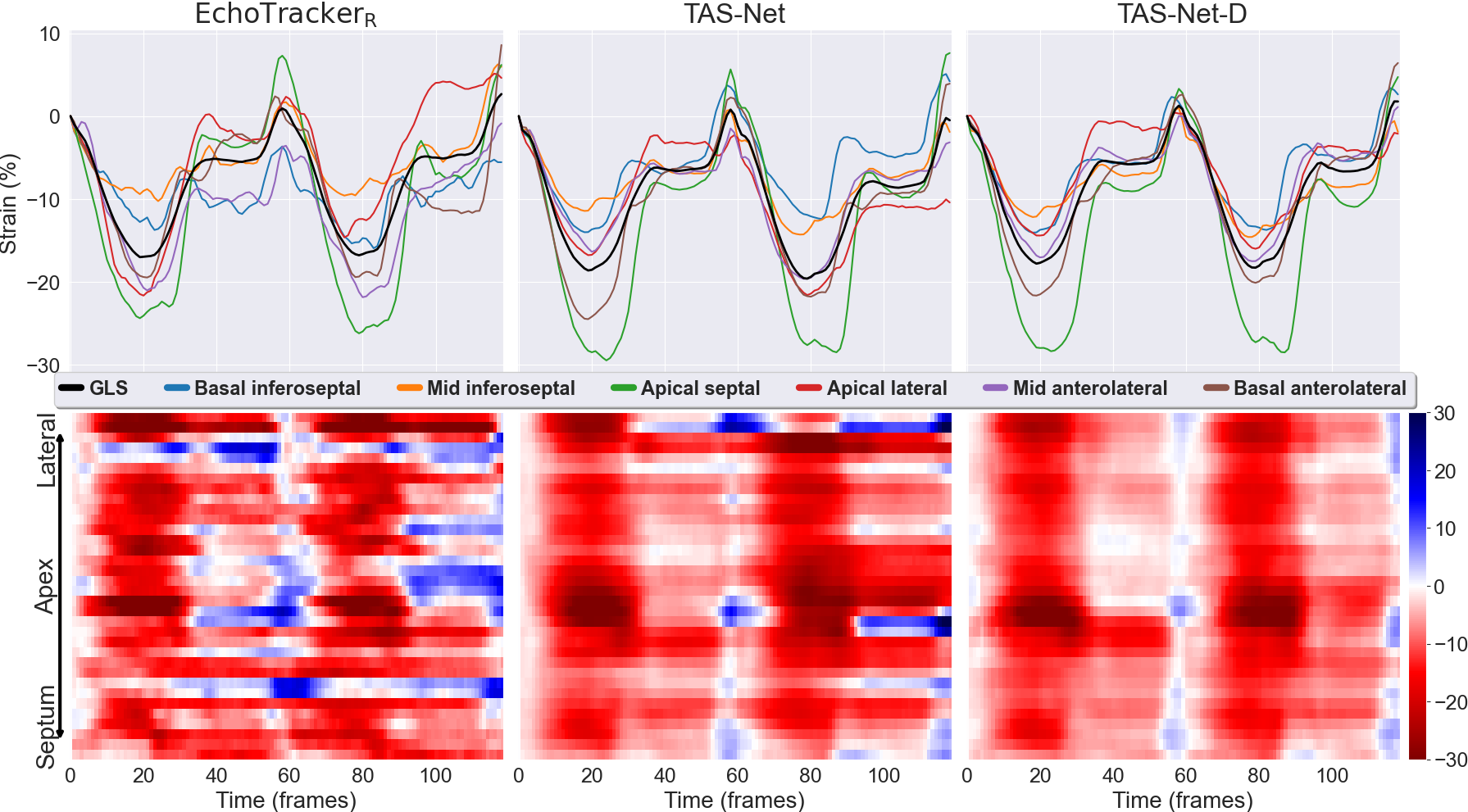}
\caption{Example of multi-cycle strain curves and maps from CAMUS. Top: GLS and 6 RLS curves. 
Bottom: 
strain maps 
at each instant and each point along the centerline.
TAS-Net-D exhibits greater coherence between the two cycles for strain curves and maps, and increased spatial consistency in strain maps, even though the latter is not explicitly optimized.}
\label{fig:multi_samples}
\end{figure*}

We evaluated the best-performing methods, $\text{EchoTracker}_\text{R}$ and TAS-Net-D, in a multi-cycle setting. We used 120 videos from the CAMUS dataset, which are in the validation set (none of these videos are used for training in the simulated or real datasets) and contain at least two cycles. We used the original CAMUS dataset masks to initialize the tracking and tracked for two or more cycles. We used each method's predicted mesh to compute the area of the LV and obtain the ED and ES instants, as there are no annotations for instants other than the first ED and ES. We computed and compared the GLS obtained from the first cycle (ED1/ES1) and the second cycle (ED2/ES2). The Bland-Altman analysis for these samples is shown in \cref{fig:ba}, complemented by the qualitative analysis in \cref{fig:multi_samples}. These figures consist of the GLS and regional strain curves as well as the strain map obtained by computing the strain locally between each neighboring point in the centerline. Considering none of these methods were trained on multi-cycle videos, they achieve respectable performance. TAS-Net-D has the lowest variance ($\sigma$) for the GLS and RLS (\cref{fig:ba}).
The strain maps show how TAS-Net-D increases the coherence between the two cycles at a local level.

\section{Conclusion}
We 
presented 
TAS-Net-D
for deep learning myocardial tracking, which 
mitigates drift by extending TAS-Net beyond its sliding-window formulation. A modified refinement scheme and window-persistent tokens let information flow across windows through fine-tuning on real echocardiographic videos, reducing drift and improving robustness in test–retest and multi-cycle evaluations. 
Drift should be minimized but 
zero drift 
may not be desired
since acquisition variability can induce apparent geometric changes; our edge-length metric is invariant to in-plane rotation and translation, though out-of-plane rotations can still affect projected distances. This work is a step toward fully automatic strain estimation, with future work integrating automatic segmentation to remove manual initialization.


\section{Acknowledgements}

This work was supported in part by the NSERC’s Discovery Grants and Canada Graduate Scholarships-Doctoral programs, the FRQNT Doctoral Training Scholarships, the French National Research Agency (LABEX PRIMES [ANR-11-LABX-0063] of Université de Lyon, and ORCHID [ANR-22-CE45-0029-01] projects) and by the Fédération Française de Cardiologie - FOCACHIA project 2025-2026.

\bibliographystyle{splncs04}
\bibliography{ref}

\end{document}